\documentclass[conference]{IEEEtran}

\usepackage{graphicx}
\usepackage{amsmath,bbm,epsfig,amssymb,amsfonts,amstext,verbatim,amsopn,cite,subfigure,multirow,multicol,lipsum}
\usepackage{balance}
\usepackage{url}
\usepackage{amsfonts}
\usepackage{epsfig}
\usepackage{setspace}
\usepackage{stmaryrd}
\usepackage{psfrag}	
\usepackage{multirow}
\usepackage{float}
\usepackage[process=auto]{pstool}
\usepackage{etoolbox}
\usepackage{algorithm}
\usepackage{algorithmic}
\allowdisplaybreaks

\newtoggle{OneColumn}

\toggletrue{OneColumn}

\iftoggle{OneColumn}{%
  
}{%
}

\EndPreamble
\begin{document}

\title{Symbol Synchronization for Diffusive Molecular Communication Systems \vspace{-0.3cm}}

\author{Vahid Jamali, Arman Ahmadzadeh, and Robert Schober \\
\IEEEauthorblockA{ Friedrich-Alexander University (FAU), Erlangen, Germany \vspace{-0.1cm}
}
}

\maketitle

\begin{abstract}
Symbol synchronization refers to the estimation of  the start of a symbol interval and is needed for reliable detection. In this paper, we develop a symbol synchronization framework for molecular communication (MC) systems where we consider some practical challenges which have not been addressed in the literature yet. In particular, we take into account that in MC systems, the transmitter may not be equipped with an internal clock and may not be able to emit molecules with a fixed release frequency. Such restrictions hold for practical nanotransmitters, e.g. modified cells, where the lengths of the symbol intervals may vary due to the inherent randomness in the availability of food and energy for molecule generation, the process for molecule production, and the release process.  To address this issue, we propose to employ two types of molecules, one for synchronization and one for data transmission. We derive the optimal maximum likelihood (ML) symbol synchronization scheme as a performance upper bound. Since ML synchronization entails high complexity,  we also propose two low-complexity synchronization schemes, namely a peak observation-based scheme and a threshold-trigger scheme, which are suitable for MC systems with limited computational capabilities. Our simulation results reveal the effectiveness of the proposed synchronization~schemes and suggest that the end-to-end performance of MC systems significantly depends on the accuracy of symbol synchronization\footnote{This paper has been accepted for presentation at IEEE ICC 2017.}. 
\end{abstract}

\section{Introduction}

Recent advances in biology, nanotechnology, and medicine have enabled the possibility of communication in nano/micrometer scale environments \cite{CellBio,Nariman_Survey,ArmanMobileMC}.  Employing molecules as information carriers, molecular communication (MC) has quickly emerged as a bio-inspired approach for man-made communication systems in such  environments. In fact, calcium signaling among neighboring cells, the use of neurotransmitters for communication across the synaptic cleft of neurons, and the exchange of autoinducers as signaling molecules in bacteria for quorum sensing are among the many examples of MC in nature \cite{CellBio}.

\subsection{Prior Work on Synchronization in MC}

One of the crucial requirements for establishing a reliable communication link is symbol synchronization where the start of a symbol interval is determined at the receiver.  Most  works available in the literature on MC assume perfect symbol synchronization for data detection, see e.g. \cite{Nariman_Survey,Arman_ReactReciever,Chae_Absorbing}. First studies for establishing a synchronization mechanism in MC have been conducted in  \cite{MC_Akyildiz_Sync,MC_Nakano_Sync,MC_Sync_Bio,MC_Clock_Sync,MC_Gaussian_Sync,MC_Blind_Sync}. In particular, in \cite{MC_Akyildiz_Sync,MC_Nakano_Sync,MC_Sync_Bio}, the authors proposed a  scheme for synchronizing multiple molecular machines as a means to enable the collaborative achievement of a common task, e.g., synchronization via quorum sensing among bacteria to coordinate their behavior. \textit{Symbol synchronization} was investigated in \cite{MC_Clock_Sync,MC_Gaussian_Sync,MC_Blind_Sync}. In \cite{MC_Clock_Sync} and \cite{MC_Gaussian_Sync}, the authors proposed a two-way message exchange mechanism between the transmitter and the receiver where \textit{constant} frequency and delay offsets between the clocks of the transmitter and the receiver  are estimated and corrected. However, to achieve high performance, the synchronization protocols in \cite{MC_Clock_Sync} and \cite{MC_Gaussian_Sync} require several rounds of two-way message exchange between the transmitter and the receiver which leads to a huge overhead considering the slow propagation of molecules in  MC channels. Futhermore, for cases when flow is present in the entvinronment e.g. in the direction from the transmitter to the receiver, it may not be possible to establish a feedback link  from the receiver to the transmitter. To reduce the synchronization overhead, the authors in \cite{MC_Blind_Sync} proposed a blind synchronization scheme based on a sequence of data molecules observed at the receiver. However, in \cite{MC_Blind_Sync}, the clocks of the transmitter and the receiver are assumed to have identical frequencies and only a \textit{constant} clock offset may exist.

\subsection{Our Contributions}

In this paper, we aim to develop a new symbol synchronization framework by taking into account some practical challenges of MC systems which have not been addressed in \cite{MC_Clock_Sync,MC_Gaussian_Sync,MC_Blind_Sync}. In particular, in \cite{MC_Clock_Sync,MC_Gaussian_Sync,MC_Blind_Sync}, similar to wireless communications \cite{Clock_Sync}, it is assumed  that the nodes are equipped with internal clocks and accurate oscillators. Thereby, the problem of synchronization was reduced to the elimination of possible frequency and delay offsets between the clocks. Furthermore, in \cite{MC_Clock_Sync,MC_Gaussian_Sync,MC_Blind_Sync}, it is assumed that the transmitter emits molecules  with a perfect release frequency, i.e., the  symbol durations are constant and identical.  However, in a real MC system, the transmitter will be a biological or electronic nanomachine, e.g. a modified cell, which controls the release of the information molecules into the channel using e.g. electrical, chemical,
or optical signals \cite{CellBio,Hamid_Ion_Modul}. Because of the inherent randomness in the availability of food and energy for molecule generation, the process for molecule production, and the release process, see \cite[Chapters~12 and 13]{CellBio},  the lengths of the symbol intervals may vary in practical MC systems. 

To cope with the aforementioned practical challenges, in this paper, we develop a symbol synchronization framework for MC where the transmitter is not necessarily equipped with an internal clock nor restricted to release the molecules with a constant frequency. To enable symbol synchronization, we employ two types of molecules, one  for synchronization  and one for data transmission. We first derive the optimal maximum likelihood (ML) symbol synchronization scheme as a performance upper bound for the proposed synchronization framework. Since ML synchronization entails high complexity, we also propose two low-complexity synchronization schemes, namely a peak observation-based (PO) scheme and a  threshold-trigger (TT) scheme, which are suitable for MC systems with limited computational capabilities. Our simulation results reveal the effectiveness of the proposed synchronization~schemes and suggest that the end-to-end performance of MC systems significantly depends on  the symbol synchronization accuracy.
 
\textit{Notations:} We use the following notations throughout this paper: $\mathbbmss{E}\{\cdot\}$ denotes expectation and $|\cdot|$ represents the cardinality of a set.   $\lceil x\rceil$ denotes the ceiling function which maps real number $x$ to the smallest integer number which is larger than or equal to  $x$. $\mathcal{P}(\lambda)$  denotes a Poisson random variable (RV) with mean  $\lambda$ and $f_{\mathcal{P}}(x,\lambda)=\frac{\lambda^xe^{-\lambda}}{x!}$ is the probability mass function (PMF) of a Poisson RV with mean $\lambda$.

\section{System and Signal Models}

In this section, we first present the  MC system model considered in this paper. Subsequently, we introduce the signal models used for synchronization and data transmission.

\subsection{System Model}

\begin{figure}
  \centering
 \scalebox{0.85}{
\pstool[width=1.1\linewidth]{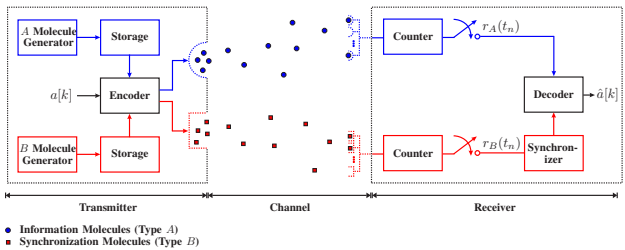}{
\psfrag{AA}[c][c][0.5]{\textbf{Generator}}
\psfrag{AB}[c][c][0.5]{$A$ \textbf{Molecule}}
\psfrag{BA}[c][c][0.5]{$B$ \textbf{Molecule}}
\psfrag{BB}[c][c][0.5]{\textbf{Storage}}
\psfrag{CC}[c][c][0.5]{\textbf{Encoder}}
\psfrag{DD}[c][l][0.6]{$a[k]$}
\psfrag{EE}[c][c][0.5]{\textbf{Counter}}
\psfrag{HH}[c][r][0.6]{$r_A(t_n)$}
\psfrag{GG}[c][r][0.6]{$r_B(t_n)$}
\psfrag{F1}[c][c][0.5]{\textbf{Synchron-}}
\psfrag{F2}[c][c][0.5]{\textbf{izer}}
\psfrag{JJ}[c][r][0.6]{$\hat{a}[k]$}
\psfrag{II}[c][c][0.5]{\textbf{Decoder}}
\psfrag{KK}[c][c][0.5]{\textbf{Transmitter}}
\psfrag{LL}[c][c][0.5]{\textbf{Channel}}
\psfrag{MM}[c][c][0.5]{\textbf{Receiver}}
\psfrag{D}[l][c][0.5]{\textbf{Information Molecules (Type $A$)}}
\psfrag{S}[l][c][0.5]{\textbf{Synchronization Molecules (Type $B$)}}
} }  \vspace{-0.01cm}
\caption{Block diagram of the considered MC setup. The  molecules released by the transmitter for information transmission and synchronization are shown as blue circles and red squares, respectively. \vspace{-0.03cm}}
\label{Fig:Block}
\end{figure}

We consider an MC system  consisting of a transmitter, a channel, and a receiver, see Fig.~\ref{Fig:Block}. The transmitter releases two types of molecules, namely type-$A$ and type-$B$ molecules, where type-$A$ molecules are used for information transmission whereas type-$B$ molecules are employed for synchronization.  The released molecules diffuse through the fluid medium between the transmitter and the receiver.  The movements of individual molecules are assumed to be independent from each other. Furthermore, we assume that molecules of types $A$ and $B$ have idential diffusion coefficients denoted by $D$ \cite{Nariman_Survey}. We consider a spherical receiver whose surface is partially covered by two different types of receptors for detecting type-$A$ and type-$B$ molecules, respectively  \cite{Arman_ReactReciever}. Molecules which reach the receiver can participate in a reversible bimolecular reaction with receiver receptor proteins. Thereby, the receiver treats the time-varying numbers of bound type-$A$ and type-$B$ molecules as the received signals for data detection and synchronization, respectively.  

The MC channel is characterized by the following two quantities. \textit{i)} The \textit{expected} number of type-$x$ molecules bound to the corresponding receptors at the receiver at time $t$ due to the release of molecules by the transmitter in \textit{one} symbol interval starting at $t=0$, which is denoted by $P_x(t),\,\,x\in\{A,B\}$. \textit{ii)} The \textit{expected} number of external noise molecules bound to the receptors at \textit{any} given time, denoted by $z_x,\,\,x\in\{A,B\}$. In general, $P_x(t),\,\,x\in\{A,B\}$,  depends on the release mechanism at the transmitter, the MC environment, and properties of the receiver such as its size, the number receptors, etc. For instance, assuming instantaneous molecule release and a point source transmitter,   expressions for $P_x(t)$ can be found in \cite{Arman_ReactReciever} for a general reactive receiver and in \cite{Chae_Absorbing} for an absorbing receiver. On the other hand, the external noise molecules originate from other MC links  or natural sources which also employ type-$A$ or type-$B$ molecules. We emphasize that the synchronization and detection schemes proposed in this paper are general and are applicable for any given expression of $P_x(t)$ and any value of $z_x$. For future reference, we refer to ${\mathtt{SNR}}_x=\frac{{\mathrm{max}}_{t\geq 0}\,\, P_x(t)}{z_x},\,\,x\in\{A,B\}$, as the signal-to-noise ratio (SNR) for type-$x$ molecules.

\subsection{Signal Model}

Let  $a[k]\in\{0,1\}$ denote a binary data symbol in the $k$-th symbol interval.  The transmitter wishes to continuously send data symbols; however, the release time of the molecules at the transmitter may vary from one symbol interval to the next due to variations in the availability of food and energy for molecule generation, the rate for molecule production, and the release process  over time, see \cite[Chapters~12 and 13]{CellBio}. To model the aforementioned effects, let $t_s[k]\in\mathcal{T}[k]$ denote an RV whose realization specifies the start of the $k$-th symbol interval where $\mathcal{T}[k]$ is given by
\begin{IEEEeqnarray}{lll} \label{Eq:ML_Tk}
\mathcal{T}[k]=[t_s[k-1]+T^{\min},t_s[k-1]+T^{\max}].
\end{IEEEeqnarray}
The duration of the $k$-th symbol interval is the time elapsed between $t_s[k]$ and $t_{s}[k+1]$; hence, in (\ref{Eq:ML_Tk}), $T^{\min}$ and $T^{\max}$ are in fact the  minimum and maximum possible lengths of a symbol interval, respectively. In other words, the length of each symbol interval is an RV in $[T^{\min},T^{\max}]$. Note that the symbol rate of the considered MC system, denoted by $R$, is bounded by $\frac{1}{T^{\max}}\leq R \leq \frac{1}{T^{\min}}$.

To establish symbol synchronization, at the beginning of each symbol interval, the transmitter releases $N_B$ type-$B$ molecules. Moreover, depending on whether $a[k]=1$ or $a[k]=0$ holds, the transmitter releases either $N_A$ or zero type-$A$ molecules, respectively, i.e., ON-OFF keying modulation is performed \cite{Nariman_Survey}\footnote{In real MC systems, the number of  molecules released by the transmitter may not be constant and may also vary from one symbol interval to the next. For simplicity, in this paper, we assume that the transmitter waits until a sufficient number of molecules is available and then releases exactly $N_B$ synchronization and/or $N_A$ information molecules, respectively. In future work, we will extend the proposed synchronization schemes  to account for varying numbers of released molecules.}. To model the received signal, we assume that the receiver periodically counts the numbers of information and synchronization molecules bound to the respective receptors on its surface with a frequency of $\Delta t$ seconds. Therefore, time can be discretized into a sequence of observation time samples $t_n=(n-1)\Delta t,\,\,n=1,2,\dots$, at the receiver. Moreover, let us define  $r_x(t_n)$ as the number of type-$x$ molecules bound to the respective receptors  at sample time $t_n$. Since at any given time after the release of the molecules by the transmitter, the molecules are either bound to a receptor or not, a binary state model applies and the number of bound molecules follows a binomial distribution. We note that the binomial distribution converges to the Poisson distribution when the number of trials is high and the success probability is small \cite{Adam_Enzyme}. These assumptions are justified for MC since the number of released molecules is typically very large
and the probability that a molecule released by the transmitter reaches the receiver is typically very small \cite{TCOM_MC_CSI}. Therefore,  $r_x(t_n)$ can be modeled as follows \cite{TCOM_MC_CSI}
\begin{IEEEeqnarray}{lll} \label{Eq:InputOutput}
  r_x(t_n)  =  \mathcal{P}(\bar{r}_x(t_n)), \quad x\in\{A,B\},
\end{IEEEeqnarray}
where $\bar{r}_x(t_n)=\mathbbmss{E}\{r_x(t_n)\}$ is given by
\begin{IEEEeqnarray}{lll} \label{Eq:ExpMol}
  \bar{r}_A(t_n)  &=  \sum_{\forall k|t_s[k]\leq t_n}  a[k] P_A\left(t_n- t_s[k]\right) + z_A, \IEEEyesnumber\IEEEyessubnumber\\
    \bar{r}_B(t_n) & =  \sum_{\forall k|t_s[k]\leq t_n}  P_B\left(t_n- t_s[k]\right) + z_B. \IEEEyessubnumber
\end{IEEEeqnarray}

\section{Proposed Synchronization Framework}

 In this section, we develop optimal and suboptimal synchronization schemes based on the communication setup introduced in Section~II. We further discuss the a priori information  about the MC channel required for the proposed synchronization schemes, the adopted detection schemes, and possible extensions of the proposed synchronization framework. 
  
\subsection{Optimal ML Synchronization}
Our goal is to determine the start of each symbol interval, i.e., $t_s[k]$, based on the received signal for type-$B$ molecules, i.e., $r_B(t_n),\,\,\forall t_n$.  \textit{Joint ML symbol} synchronization of several consecutive symbol intervals is very complicated due to the multi-dimensional nature of the corresponding ML hypothesis test. Therefore, the main challenge on which we focus here is to formulate an ML problem for \textit{symbol-by-symbol} synchronization which is numerically tractable. To this end, we introduce two assumptions which enable us to formulate an ML problem for estimating $t_s[k]$ without knowledge of $t_s[k'],\,\,k'>k$. Before presenting these assumptions, let us first define $T^{\mathrm{ow}}$ as the size of the observation window  used to compute the ML metric for each hypothesis time $t$ for $t_s[k]$, i.e.,  observation samples $t_n \in[t, t+T^{\mathrm{ow}}]$ are used for hypothesis test $t$.
 
\begin{figure}
  \centering
 \scalebox{0.8}{
\pstool[width=1\linewidth]{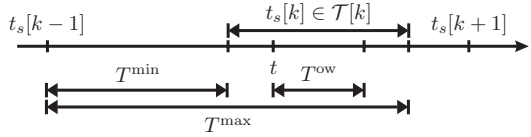}{
\psfrag{t1}[c][c][1]{$t_s[k-1]$}
\psfrag{ts}[c][c][1]{$t$}
\psfrag{t2}[c][c][1]{$t_s[k+1]$}
\psfrag{tmax}[c][c][1]{$T^{\max}$}
\psfrag{tmin}[c][c][1]{$T^{\min}$}
\psfrag{to}[c][c][1]{$T^{\mathrm{ow}}$}
\psfrag{tk}[c][c][1]{$t_s[k]\in\mathcal{T}[k]$}
} } \vspace{-0.3cm} 
\caption{Illustration of assumptions A1 and A2 adopted for the development of the symbol-by-symbol ML synchronization problem. \vspace{-0.2cm}}
\label{Fig:MLProb}
\end{figure}


\begin{itemize}
\item[A1:] We assume that $T^{\mathrm{ow}}\leq T^{\min}$ holds which ensures that the ML metric for the correct $t$, i.e., $t=t_s[k]$, is not influenced by the value of $t_s[k+1]$. 
\item[A2:] We assume that $t_s[k+1]\notin\mathcal{T}[k]$ holds which leads to the condition $T^{\max}\leq 2T^{\min}$. We note that if $t_s[k+1]\in\mathcal{T}[k]$ can  occur, $t=t_s[k+1]$ may be selected as the ML estimate for the $k$-th symbol interval.
\end{itemize}

The above assumptions are schematically illustrated in Fig.~\ref{Fig:MLProb}. Based on assumptions A1 and A2, the ML problem can be mathematically formulated as 
\begin{IEEEeqnarray}{lll} \label{Eq:ML_Sync}
\hat{t}_s^{\mathtt{ML}}[k] &= \underset{\forall t \in\mathcal{T}[k] }{\mathrm{argmax}}\,\,\prod_{t_n=t}^{t+T^{\mathrm{ow}}} f_{\mathcal{P}}\big(r_B(t_n), \bar{r}_B(t_n)|{t_s[k]=t}\big) \nonumber \\
&\triangleq\underset{\forall t \in\mathcal{T}[k] }{\mathrm{argmax}}\,\, \Lambda_B^{\mathtt{ML}}(t).\quad\,
\end{IEEEeqnarray}
 In (\ref{Eq:ML_Sync}), it is assumed that the observations  $r_B(t_n)$ at different time instances are independent such that the likelihood function over observation window $t_n\in[t,t+T^{\mathrm{ow}}]$ can be factorized into the likelihood functions for each time instance $t_n$, i.e., $f_{\mathcal{P}}\big(r_B(t_n), \bar{r}_B(t_n)|{t_s[k]=t}\big)$. Moreover, for a given hypothesis $t$ for $t_s[k]$, assuming that the  ML estimate of symbol interval $k'<k$ was correct,   $\bar{r}_B(t_n)$ is given by (\ref{Eq:ExpMol}).

\begin{figure}
  \centering
\resizebox{1\linewidth}{!}{\psfragfig{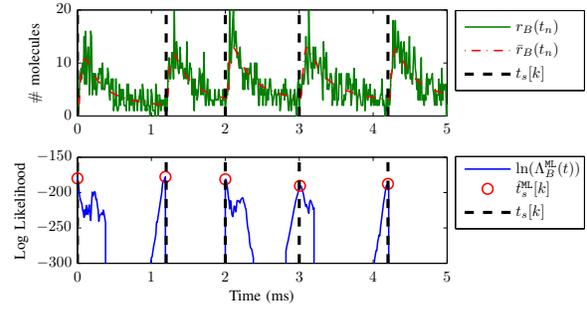}} \vspace{-0.7cm} 
\caption{Illustration of the proposed ML synchronization scheme. The details of the adopted simulation setup and the corresponding simulation parameters are given~in~Section~IV and Table~II, respectively.\vspace{-0.03cm}}
\label{Fig:Concept_ML}
\end{figure}

 Maximizing $\Lambda_B^{\mathtt{ML}}(t)$ is equivalent to maximizing $\mathrm{ln}(\Lambda_B^{\mathtt{ML}}(t))$ since $\mathrm{ln}(\cdot)$ is a monotonically increasing function. Hence, the ML problem in (\ref{Eq:ML_Sync}) can be rewritten as
\begin{IEEEeqnarray}{lll} \label{Eq:ML_log}
\hat{t}_s^{\mathtt{ML}}[k] = \underset{\forall t \in\mathcal{T}[k] }{\mathrm{argmax}}\,\,\mathrm{ln}(\Lambda_B^{\mathtt{ML}}(t))  \\
=\underset{\forall t \in\mathcal{T}[k] }{\mathrm{argmax}}\sum_{t_n=t}^{t+T^{\mathrm{ow}}} \big[ r_B(t_n)\mathrm{ln}(\bar{r}_B(t_n))-\bar{r}_B(t_n)-\mathrm{ln}(r_B(t_n)!) \big].\nonumber
\end{IEEEeqnarray}
Although the problem in  (\ref{Eq:ML_log}) does not lend itself to a simple elegant closed-form solution, we can still find the optimal ML solution numerically using a one-dimensional search and employ it as a benchmark scheme to evaluate the performance of the proposed suboptimal protocols, cf. Section~IV. Fig.~\ref{Fig:Concept_ML} illustrates an example scenario for ML synchronization for five consecutive symbol intervals. In Fig.~\ref{Fig:Concept_ML}, we choose the starts of the symbol intervals as $t_s[k]=[0, 1.2, 2, 3, 4.2]$ ms, i.e., the transmitter does not release the molecules at a fixed frequency. The proposed ML synchronization scheme is able to efficiently determine the starts of the symbol intervals for the  set of parameters considered in  Fig.~\ref{Fig:Concept_ML}.

 \subsection{Suboptimal Low-Complexity Synchronization}
The proposed ML synchronization scheme provides optimal symbol synchronization at the cost of a high computational complexity which may not be affordable for implementation at nanoscale.   Therefore, in this subsection, we propose two suboptimal low-complexity synchronization schemes which may be preferable for implementation in simple nanoreceivers.

 \subsubsection{Peak Observation-based  Synchronization}
\begin{figure}
  \centering
\resizebox{1\linewidth}{!}{\psfragfig{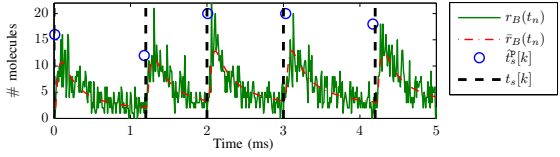}} \vspace{-0.6cm} 
\caption{Illustration of the proposed PO synchronization scheme. The details of the adopted simulation setup and the corresponding simulation parameters are given~in~Section~IV and Table~II, respectively. \vspace{-0.03cm} }
\label{Fig:Concept_Peak}
\end{figure}

 Recall that the ML synchronization scheme optimally takes into account all samples within the observation window $t_n\in[t,t+T^{\mathrm{ow}}]$ for each  possible hypothesis $t\in\mathcal{T}[k]$ in order to estimate $t_s[k]$. To reduce the complexity of ML synchronization, we propose to estimate $t_s[k]$ based on only the peak observation. To formally present the proposed PO synchronization scheme, let us first define constant $t^{\mathtt{p}}={\mathrm{argmax}}_{t\geq 0}\,\,P_B(t)$. Thereby, the set of expected time instances where the peak observation for the synchronization molecules in symbol interval $k$ can occur is given by 
\begin{IEEEeqnarray}{lll} \label{Eq:Peak_Tp}
\mathcal{T}^{\mathtt{p}}[k]=[t_s[k-1]+T^{\min}+t^{\mathtt{p}},t_s[k-1]+T^{\max}+t^{\mathtt{p}}].\quad
\end{IEEEeqnarray}
Hereby, we propose a PO symbol synchronization scheme which estimates the start of the symbol intervals as follows
\begin{IEEEeqnarray}{lll} \label{Eq:PeakProt}
\hat{t}^{\mathtt{p}}_s[k]= \Big[\underset{t_n\in \mathcal{T}^{\mathtt{p}}[k]}{\mathrm{argmax}} \,\,r_B(t_n) \Big] - t^{\mathtt{p}}.\quad
\end{IEEEeqnarray}
In Fig.~\ref{Fig:Concept_Peak}, the above PO synchronization scheme is schematically illustrated for the same example as considered in Fig.~\ref{Fig:Concept_ML}. While the complexity of the PO synchronization scheme is considerably lower than that of the ML synchronization scheme, as will be shown in detail in Section~IV, the corresponding performance loss may also be significant.  This motivates us to propose a TT synchronization scheme which is also relatively simple, but provides a better performance than the PO synchronization scheme.

 \subsubsection{Threshold-Trigger Synchronization}

 In nature, a common strategy among living organisms in response to external stimuli is based on a threshold-trigger mechanism. For example, an increase of the concentration of a specific type of molecule around a cell can trigger a corresponding response inside the cell \cite{CellBio}. In the following, we exploit the TT mechanism for symbol synchronization.

The main idea behind our simple TT symbol synchronization scheme is that the receiver considers the number of bound information molecules for detection only during the interval where the number of bound synchronization molecules is above a certain threshold. In other words, instead of determining the actual symbol interval, the proposed protocol only determines a detection zone which is used for data detection in each symbol interval.  In order to formally present the proposed scheme, let us define $\xi_B$ as a constant threshold and $\hat{t}_s^{\mathtt{thr}}[k]$ and $\hat{t}_e^{\mathtt{thr}}[k]$ as the beginning and the end of the detection zone for symbol interval $k$, respectively.  Furthermore, since the number of bound molecules is an RV and may rapidly fluctuate, we assume a minimum detection interval size of $T^{\mathrm{dw}}$ to avoid possible false alarms indicating a new symbol interval. On the other hand, $T^{\mathrm{dw}}\leq T^{\min}$ has to hold to avoid missing the next symbol interval.
In particular, the receiver determines  $\hat{t}_s^{\mathtt{thr}}[k]$ and $\hat{t}_e^{\mathtt{thr}}[k]$ as follows
\begin{IEEEeqnarray}{lll} \label{Eq:SubOpt_Sync}
\hat{t}_s^{\mathtt{thr}}[k]&=\underset{t_n>\hat{t}_e^{\mathtt{thr}}[k-1]}{\mathrm{min}}\,\,t_n|r_B(t_n)\geq \xi_B  \IEEEyesnumber\IEEEyessubnumber \\
\hat{t}_e^{\mathtt{thr}}[k]&=\mathrm{max}\Big\{\underset{t_n>\hat{t}_s^{\mathtt{thr}}[k]}{\mathrm{min}} t_n| r_B(t_n)\leq \xi_B, \hat{t}_s^{\mathtt{thr}}[k]+T^{\mathrm{dw}} \Big\},\,\,\quad \IEEEyessubnumber
\end{IEEEeqnarray}
respectively. In other words, $\hat{t}_s^{\mathtt{thr}}[k]$ in (\ref{Eq:SubOpt_Sync}a) activates detection whereas $\hat{t}_e^{\mathtt{thr}}[k]$ in (\ref{Eq:SubOpt_Sync}b) terminates detection for symbol interval~$k$. 


\begin{figure}
  \centering\vspace{-0.5cm} 
\resizebox{1\linewidth}{!}{\psfragfig{}} \vspace{-0.9cm} 
\caption{Illustration of the proposed TT synchronization scheme for $\xi_B=10$ and $T^{\mathrm{dw}}=0.8$ ms. The details of the adopted simulation setup and the corresponding simulation parameters are given~in~Section~IV and Table~II, respectively.\vspace{-0.3cm} }
\label{Fig:Concept_Bio}
\end{figure}

To further illustrate the proposed TT synchronization scheme, in Fig.~\ref{Fig:Concept_Bio}, we show  the transmission of five consecutive symbols $a[k]=1,1,0,0,1$, for $k=1,2,\dots,5$, respectively, as an example.  As can be seen, the proposed TT scheme selects many of observation samples containing information molecules within a given symbol interval for data detection without directly estimating the starts of the symbol intervals.

\subsection{Required A Priori Knowledge and Constraints}
Table~I summarizes the required a priori knowledge and underlying constraints for the proposed symbol synchronization schemes. For the considered MC system, the channel is characterized by $P_x(t)$  and $z_x$. Hence, the ML synchronization scheme requires full knowledge of the MC channel characteristics regarding the synchronization molecules. In contrast, the proposed PO and TT synchronization schemes need much less a priori information about the channel.  Nevertheless, the parameters shown in Table~I are constant for the coherence time of the MC channel. Hence, the receiver can obtain them offline at the beginning of transmission and use them for online symbol synchronization as long as the MC channel statistics remain unchanged.  We further note  that unlike the ML synchronization scheme for which the strict constraint $T^{\max}\leq 2T^{\min}$ has to hold, the proposed PO and TT synchronization schemes do not require this constraint.

\begin{table}
\label{Table:Knowledge}
\caption{Required A Priori Knowledge and Underlying Constraints for the Proposed Symbol Synchronization Schemes.} 
\begin{center}
\scalebox{0.7} { 
\begin{tabular}{|| c | c | c ||}
  \hline
 Sync. Scheme &  A Priori Knowledge & Constraints \\ \hline \hline 
 ML Sync. & $P_B(t)$  and $z_B$   &  $T^{\mathrm{ow}}\leq T^{\min}$ and $T^{\max}\leq 2T^{\min}$  \\ \hline
 PO Sync. & $t^{\mathtt{p}}$ & $-$ \\ \hline
 TT Sync. & $\xi_B$ and $T^{\mathrm{dw}}$    &  $T^{\mathrm{dw}}\leq T^{\min}$   \\ \hline 
\end{tabular}
}
\end{center}
\vspace{-0.3cm}
\end{table}

\subsection{Detection} 

In this paper, we adopt two simple  detectors to evaluate the bit error rate (BER) achieved with the proposed synchronization schemes. In particular, we use the following threshold detectors employing the mean and the peak numbers of information molecules observed in each symbol interval for detection \cite{Adam_OptReciever}, respectively, 
\begin{IEEEeqnarray}{lll} 
\hat{a}^{\mathtt{mean}}[k] =\begin{cases}
1,\quad &\mathrm{if}\,\,\Big[\frac{1}{N[k]}\sum_{t_n=\hat{t}_s[k]}^{\hat{t}_e[k]} r_A(t_n) \Big]\geq \xi_A \\
0,\quad &\mathrm{otherwise}
\end{cases} \label{Eq:Mean_Det}\\
\hat{a}^{\mathtt{peak}}[k] =\begin{cases}
1,\quad &\mathrm{if}\,\, \Big[\underset{t_n\in[\hat{t}_s[k],\hat{t}_e[k]]}{\max} r_A(t_n)\Big] \geq \xi_A \\
0,\quad &\mathrm{otherwise}
\end{cases}\label{Eq:Peak_Det}
\end{IEEEeqnarray}
where $N[k]=1+\frac{\hat{t}_e[k]-\hat{t}_s[k]}{\Delta t}$ is the number of samples used for detection of the $k$-th symbol and $\xi_A$ is a constant threshold.  $\hat{t}_s[k]$ and $\hat{t}_e[k]$ are the start and the end of the detection interval for symbol interval $k$, and depending on the adopted synchronization scheme, are given by
\begin{IEEEeqnarray}{lll}\label{Eq:Ts_tot} 
(\hat{t}_s[k],\hat{t}_e[k])   = \begin{cases}
(\hat{t}_s^{\mathtt{ML}}[k],\hat{t}_s^{\mathtt{ML}}[k+1]),&\text{ML Sync.} \\
(\hat{t}_s^{\mathtt{p}}[k],\hat{t}_s^{\mathtt{p}}[k+1]),&\text{PO Sync.} \\
(\hat{t}_s^{\mathtt{thr}}[k],\hat{t}_e^{\mathtt{thr}}[k]),&\text{TT Sync.}
\end{cases}\quad\,\,
\end{IEEEeqnarray}
 We note that the mean detector in (\ref{Eq:Mean_Det}) leads to a lower BER than the peak detector in (\ref{Eq:Peak_Det}). However, the latter has a lower complexity as it employs only one sample observation for detection. Hence, these two detectors enable different tradeoffs between performance and complexity, see \cite{Adam_OptReciever} and Section~IV.

\subsection{Extensions}

In the following, we present two possible extensions of the synchronization framework developed in this paper.

\textit{Extension~1:} In this paper, we consider a point-to-point MC system. However, the proposed synchronization framework is also applicable  for the broadcast channel, i.e., when a transmitter wishes to communicate with multiple receivers. In this case, the transmitter may employ different types of information molecules for each receiver, e.g., type $A_1,A_2,\dots$, and $A_M$ molecules for receivers $1,2,\dots$, and $M$, respectively. However, in such a broadcast channel, only one type of synchronization molecule, e.g., type $B$, is sufficient to synchronize all links, provided that the transmitter employs the same symbol interval for all types of emitted molecules. Hence, each receiver can independently apply the synchronization and detection schemes presented in this paper. An advantage of the proposed synchronization framework is that  as the number of receivers increases, the total synchronization overhead (in terms of the required resources for synchronization) remains constant.

\textit{Extension~2:} 
A common challenge of imperfect symbol synchronization are deletion and insertion errors \cite{InsDelMarkerCode,InsDelCode}. A deletion error occurs if the adopted synchronization protocol fails to identify the start of a symbol interval, and an insertion error occurs if a false alarm introduces an additional symbol interval. To cope with this challenge in conventional communication systems, special codes were designed which are capable of  correcting a codeword corrupted by insertions and deletions \cite{InsDelMarkerCode,InsDelCode}. Therefore, it is interesting to investigate which deletion/insertion codes are effective in combination with the MC synchronization framework developed in this paper and to potentially develop new codes specifically for MC systems. However, addressing these questions is beyond the scope of this paper but constitutes an interesting  direction for future work.

\section{Simulation Results}

\begin{table}
\label{Table:Parameter}
\caption{Default Values for Simulation Parameters \cite{Arman_ReactReciever,NanoCOM16}.\vspace{-0.2cm}}  
\begin{center}
\scalebox{0.52} { 
\begin{tabular}{|| c | c || c | c ||}
  \hline
 Parameter &  Definition &  Value \\ \hline \hline 
$N_A,N_B$ & Number of type-$A$ and -$B$  released molecules  & $10^3$ molecules  \\  \hline
$n_A,n_B$ & Number of type-$A$ and -$B$ receptors & $10^3$ receptors   \\  \hline
$D$ & Diffusion coefficient of type-$A$ and -$B$ molecules & $5\times 10^{-9}$ $\text{m}^2\cdot\text{s}^{-1}$   \\  \hline
$r_0$ & Distance between the transmitter and the receiver & $2$ $\mu$m \\  \hline
$r_r$ & Radius of the spherical receiver & $1$ $\mu$m   \\  \hline
$k_f$ & Forward reaction rate for molecule binding& $25\times 10^{-14}$  $\text{m}^3\cdot\text{moleclue}^{-1}\cdot\text{s}^{-1}$  \\  \hline
$k_r$ & Backward reaction rate for molecule binding & $5\times 10^{4}$ $\text{s}^{-1}$   \\  \hline   
$\Delta t$ & Sampling time at the receiver & $10$ $\mu$s   \\  \hline  
$T^{\mathrm{min}}$ & Minimum length of a symbol time & $0.8$ ms   \\  \hline          
$T^{\mathrm{max}}$ & Maximum length of a symbol time & $1.2$ ms   \\  \hline   
$T^{\mathrm{ow}}$ & Length of the ML observation window & $T^{\mathrm{min}}$  \\  \hline    
$T^{\mathrm{dw}}$ & Length of the detection window for threshold-trigger sync. & $T^{\mathrm{min}}$  \\  \hline     
$\mathtt{SNR}_A,\mathtt{SNR}_B$ & SNR for type-$A$ and -$B$  molecules & $10$ dB   \\  \hline            
\end{tabular}
}
\end{center}
\vspace{-0.3cm}
\end{table}

\begin{figure*}[!tbp]
  \centering
  \begin{minipage}[b]{0.3\textwidth}
  \centering
    \resizebox{1.1\linewidth}{!}{
\psfragfig{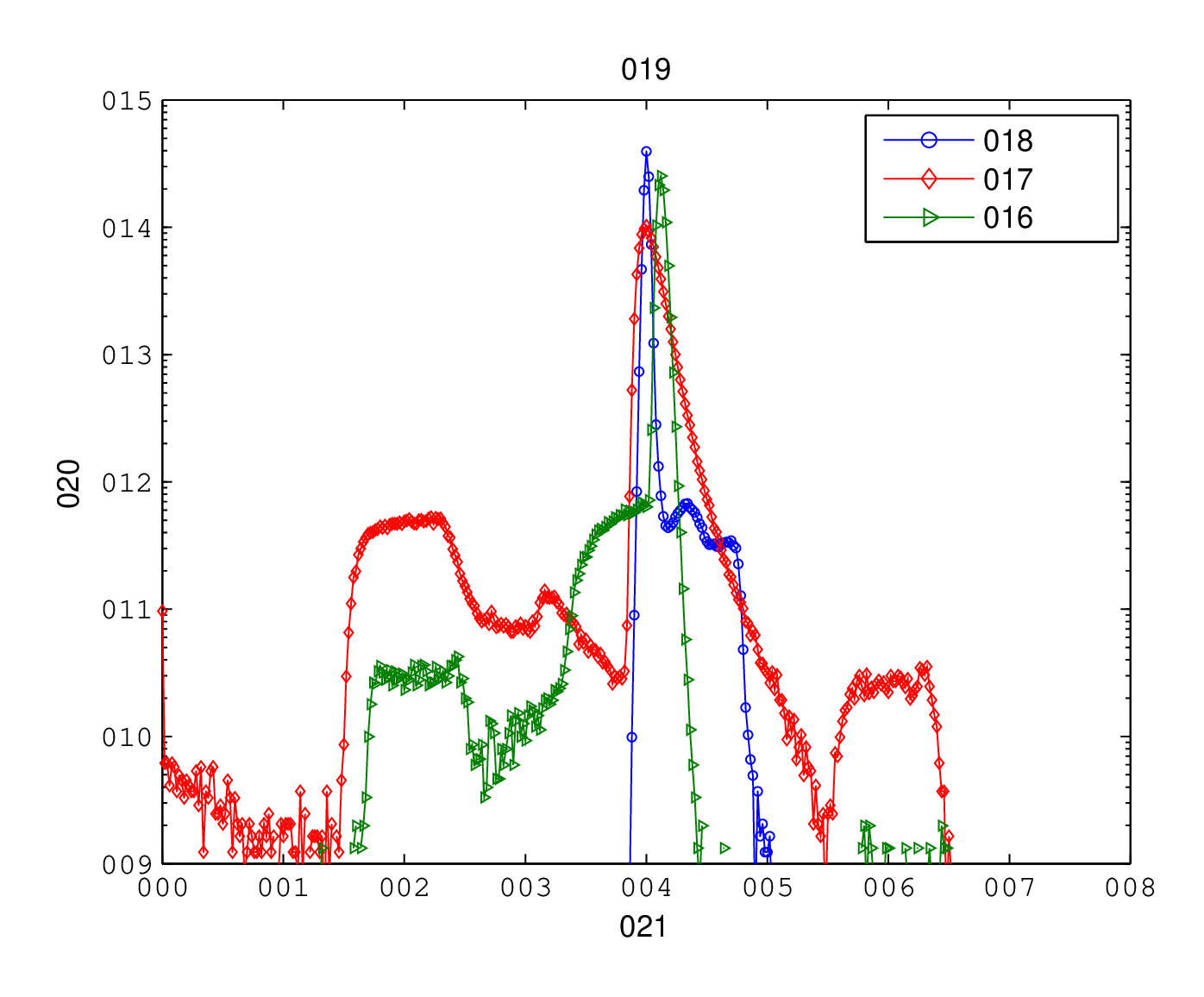}}
  \end{minipage}
  \hfill
  \begin{minipage}[b]{0.3\textwidth}
  \centering
    \resizebox{1.1\linewidth}{!}{
\psfragfig{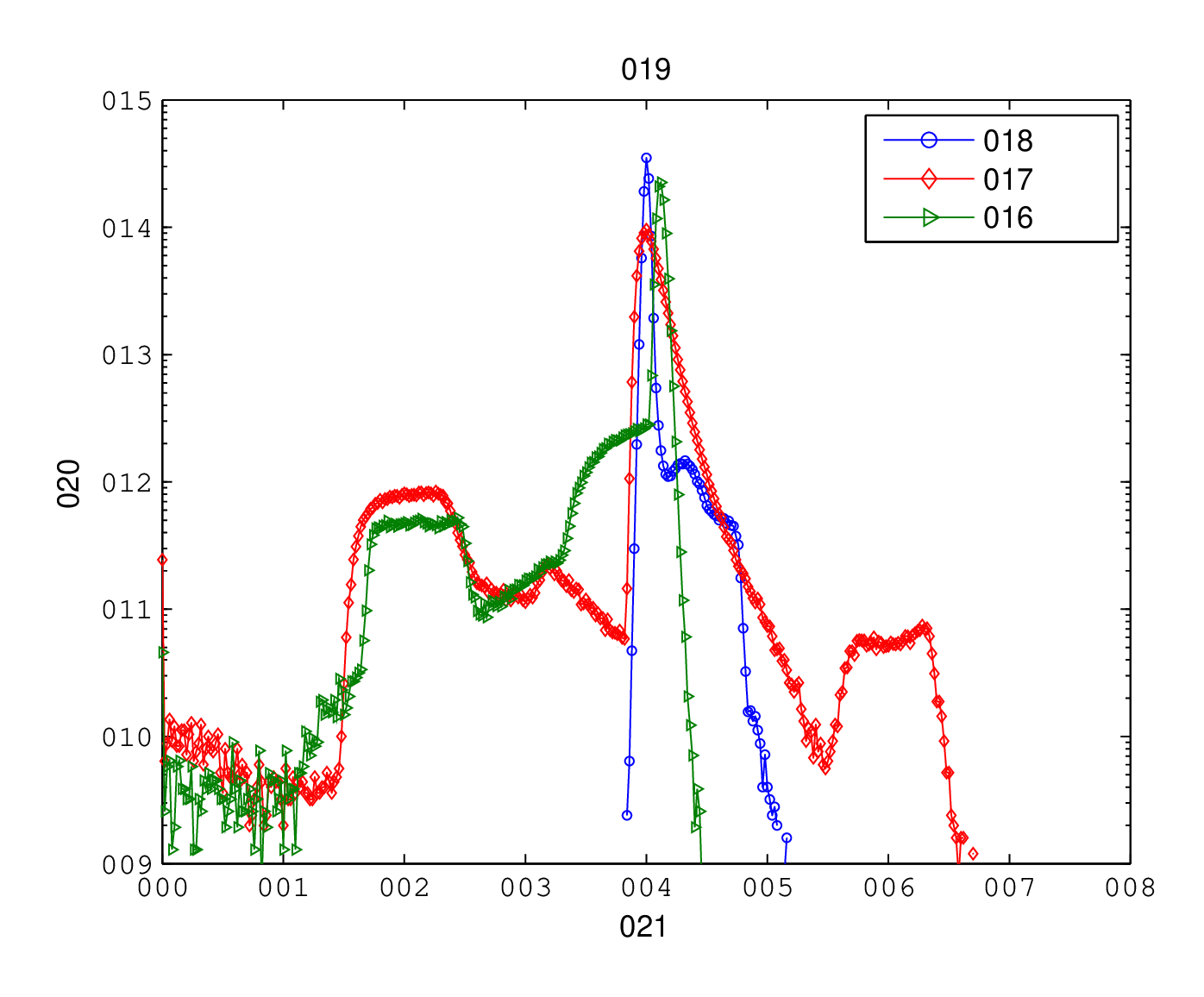}}
  \end{minipage}
    \hfill
  \begin{minipage}[b]{0.3\textwidth}
  \centering
    \resizebox{1.1\linewidth}{!}{
\psfragfig{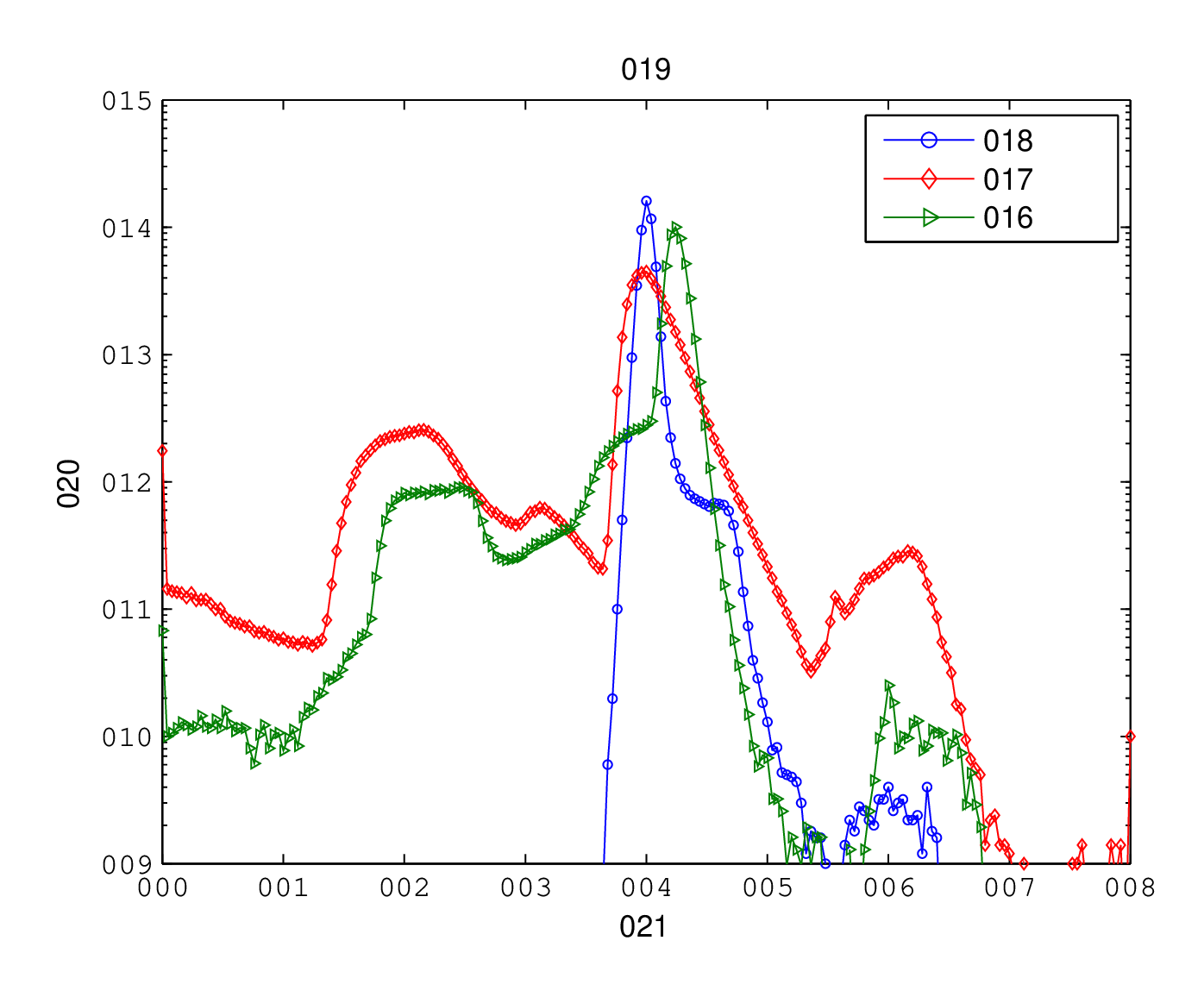}}
  \end{minipage}
      \hfill
  \begin{minipage}[b]{0.1\textwidth}
  \end{minipage}\vspace{-0.4cm}
\caption{ Estimated PDF (histogram) of the normalized synchronization error, $\bar{e}_t[k]$ for a) $\mathtt{SNR}_B=10$ dB and  $[T^{\min},T^{\max}]=[0.8, 1.2]$ ms ($\bar{T}^{\mathrm{symb}}=1$ ms), b) $\mathtt{SNR}_B=5$ dB and  $[T^{\min},T^{\max}]=[0.8, 1.2]$ ms ($\bar{T}^{\mathrm{symb}}=1$ ms), and c) $\mathtt{SNR}_B=10$ dB and  $[T^{\min},T^{\max}]=[0.4, 0.6]$ ms ($\bar{T}^{\mathrm{symb}}=0.5$ ms). For the TT synchronization scheme, the synchronization threshold is chosen as $\xi_B=13,15$, and $17$ for the results shown in a), b), and c), respectively.}
\label{Fig:PDF}  
\end{figure*}

In this section, we provide simulation results to evaluate the effectiveness of the proposed synchronization schemes. For simplicity, we assume instantaneous molecule release and a point source transmitter and employ the reactive receiver model recently developed in \cite{Arman_ReactReciever} for the calculation of $P_x(t),\,\,x\in\{A,B\}$. Moreover, we assume that $t_s[k]$ is uniformly distributed in $\mathcal{T}[k]$, i.e.,  the length of each symbol interval is an RV uniformly distributed in the interval $[T^{\min},T^{\max}]$. Furthermore, we consider blocks of $K=50$ symbol intervals. For the perfect, OP, and TT synchronization schemes, we average our results over $10^5$ blocks whereas for the ML synchronization scheme, we  average our results over $2\times 10^3$ blocks due to the high computational complexity.  Unless otherwise stated, we adopt the default values of the system parameters given in Table~II. In order to compare the performances of the considered synchronization schemes, we define the normalized synchronization error as  
\begin{IEEEeqnarray}{lll} 
\bar{e}_t[k] = \frac{\hat{t}_s[k]-t_s[k]}{\bar{T}^{\mathrm{symb}}},
\end{IEEEeqnarray}
where $\hat{t}_s[k]$ is given in (\ref{Eq:Ts_tot}) for the three proposed synchronization schemes and  $\bar{T}^{\mathrm{symb}}$ is the average symbol duration, i.e., $\bar{T}^{\mathrm{symb}}=\frac{T^{\max}+T^{\min}}{2}$.

 In Fig.~\ref{Fig:PDF}, we show the histogram of $\bar{e}_t[k]$ and we highlight some interesting observations  from this figure in the following.  First, we observe that the peaks of the probability density function (PDF) for the ML and PO synchronization schemes are centered at $\bar{e}_t[k]=0$ whereas for the TT synchronization scheme, the peak of the PDF is at a positive value of $\bar{e}_t[k]$. This is expected since the TT synchronization scheme does not aim to estimate the start of the symbol intervals and only determines a detection zone within each symbol interval. Fig.~\ref{Fig:PDF} reveals also the presence of insertion and deletion errors for the proposed synchronization schemes, cf. Subsection~III.E. In particular,  small values of $|\bar{e}_t[k]|$ correspond to no deletion and insertion errors, whereas very large and very small values of $\bar{e}_t[k]$ (i.e., $\bar{e}_t[k]>1$ and $\bar{e}_t[k]<-1$) correspond to deletion and insertion errors,  respectively. From Fig.~\ref{Fig:PDF} a), we observe that $|\bar{e}_t[k]|<0.5$ holds for the ML synchronization scheme which suggests that  deletion and insertion errors do not occur. For the proposed suboptimal schemes, we see from Fig.~\ref{Fig:PDF} a) that deletion and insertion errors are more likely for the PO synchronization scheme than for the  TT synchronization scheme since the probability that large values of $|\bar{e}_t[k]|$ occur is higher for the PO scheme than for the  TT scheme. Furthermore, we observe from Fig.~\ref{Fig:PDF} that the histograms are not symmetric with respect to $\bar{e}_t[k]=0$. This is partially due to the fact that $P_B(t)$ is not symmetric which leads to different probabilities for positive and negative values of $\bar{e}_t[k]$.  Therefore, the probabilities of insertion and deletion errors are not equal for the proposed synchronization schemes.
 
 In Fig.~\ref{Fig:PDF} b), we decrease the SNR for the synchronization molecules, i.e., $\mathtt{SNR}_B=5$ dB, compared with Fig.~\ref{Fig:PDF} a), and  keep the average symbol time unchanged, i.e., $\bar{T}^{\mathrm{symb}}=1$ ms,  whereas in Fig.~\ref{Fig:PDF} c), we keep the SNR unchanged, i.e., $\mathtt{SNR}_B=10$ dB, and use a smaller average symbol duration, i.e., $\bar{T}^{\mathrm{symb}}=0.5$ ms. We note that a smaller symbol duration leads to more inter-symbol interference (ISI). We can see that the probabilities of large values of $|\bar{e}_t[k]|$ are higher in Fig.~\ref{Fig:PDF} b) and Fig.~\ref{Fig:PDF} c) compared to Fig.~\ref{Fig:PDF} a) which leads to a poorer synchronization performance in general and more deletion and insertion errors in particular.   Moreover, the performance degradation is more severe in Fig.~\ref{Fig:PDF} c), i.e., for more ISI, than in Fig.~\ref{Fig:PDF} b), i.e., for more noise. 

\begin{figure*}[!tbp]
  \centering
  \begin{minipage}[b]{0.3\textwidth}
  \hspace{-0.5cm}
    \resizebox{1.1\linewidth}{!}{
\psfragfig{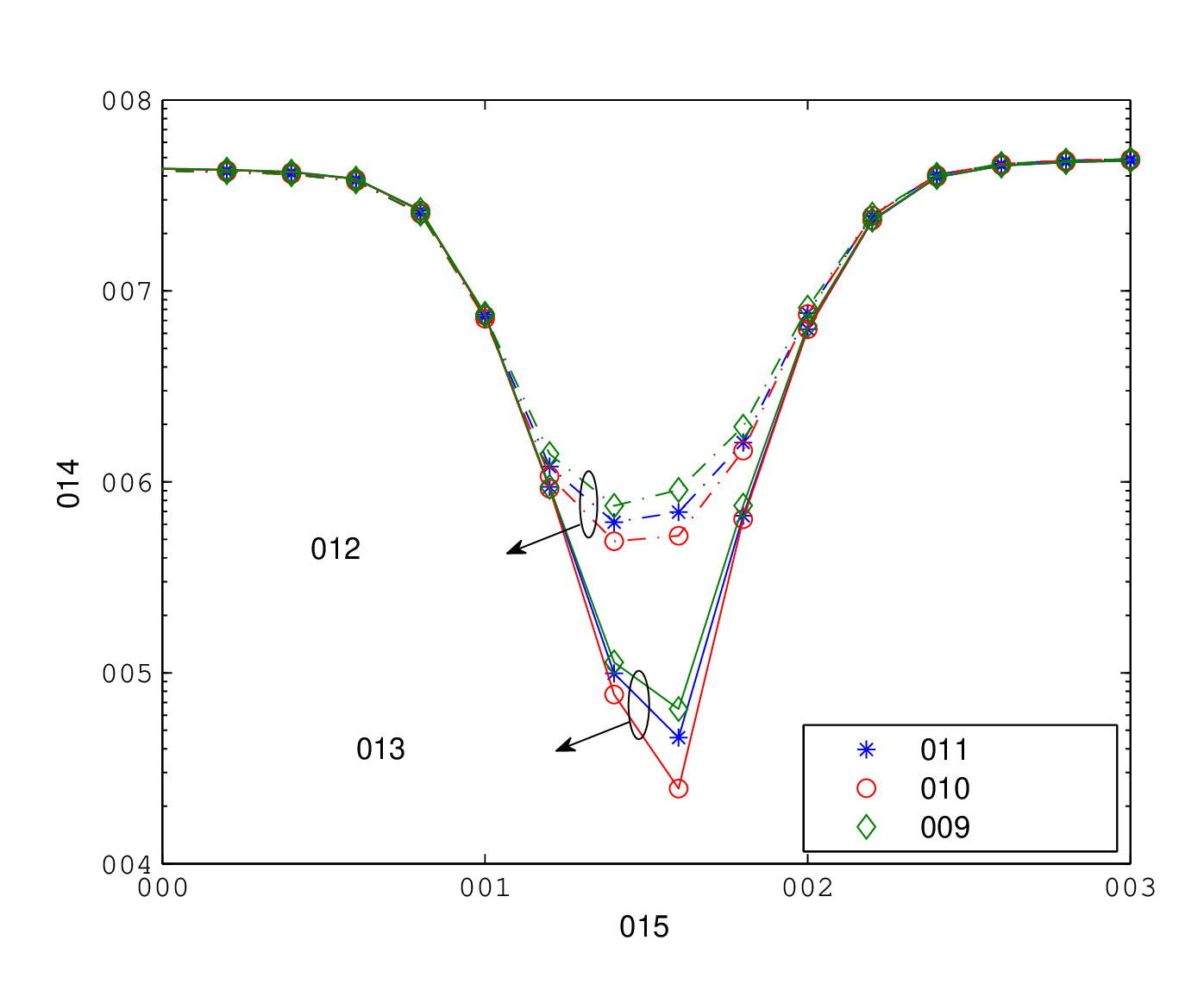}}\vspace{-0.4cm}
\caption{BER versus synchronization threshold $\xi_B$ of the TT synchronization scheme for the mean and peak detectors and different detection thresholds $\boldsymbol{\xi}_A=[\xi_A^{\mathtt{mean}},\xi_A^{\mathtt{peak}}]$. }
\label{Fig:BER_xiB}
  \end{minipage}
    \hfill
  \begin{minipage}[b]{0.3\textwidth}
  \hspace{-0.5cm}
    \resizebox{1.1\linewidth}{!}{
\psfragfig{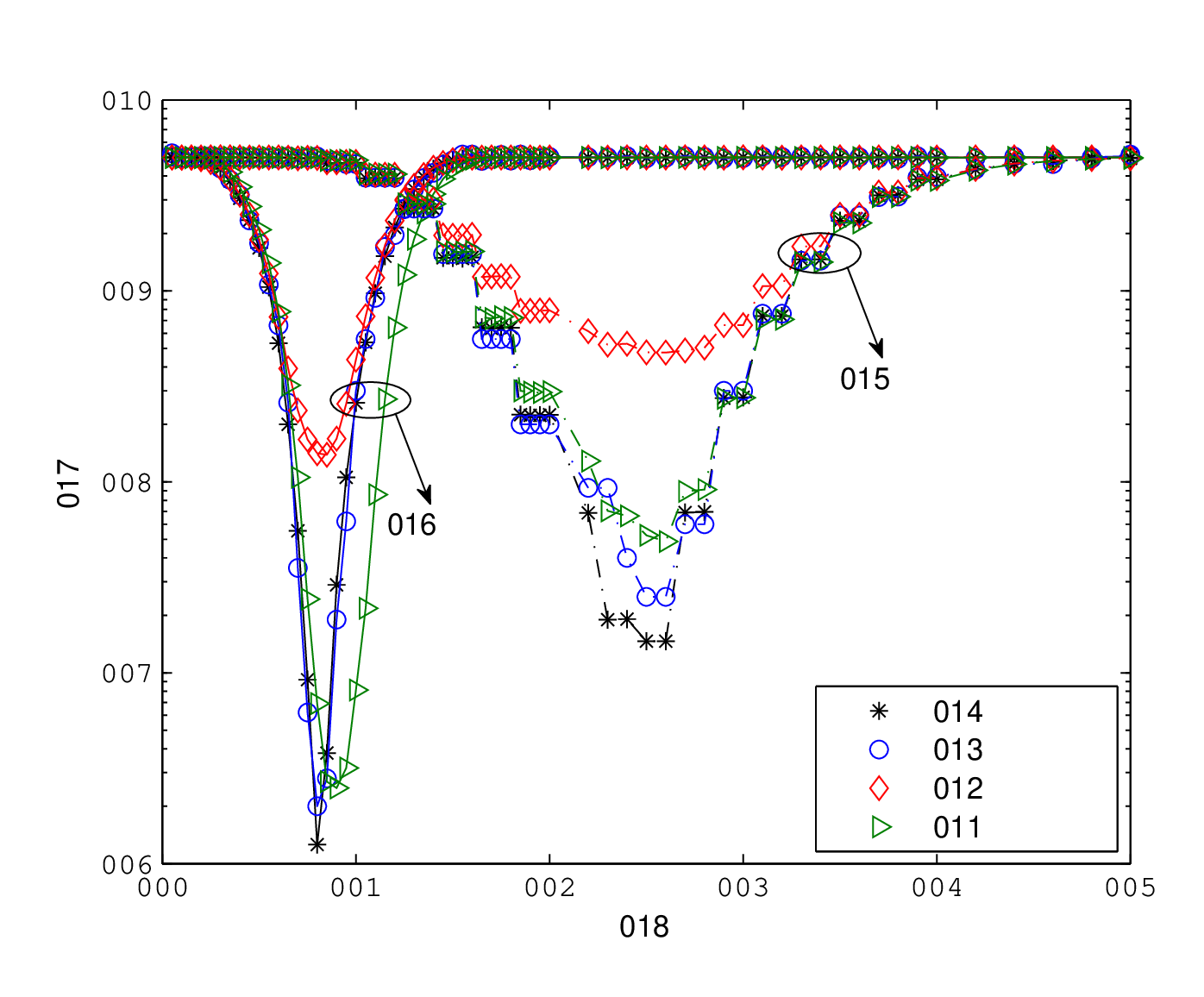}}\vspace{-0.4cm}
\caption{BER versus detection threshold $\xi_A$ for the mean and peak detectors.  }
\label{Fig:BER_xiA}
  \end{minipage}
      \hfill
  \begin{minipage}[b]{0.3\textwidth}
  \hspace{-0.5cm}
    \resizebox{1.1\linewidth}{!}{
\psfragfig{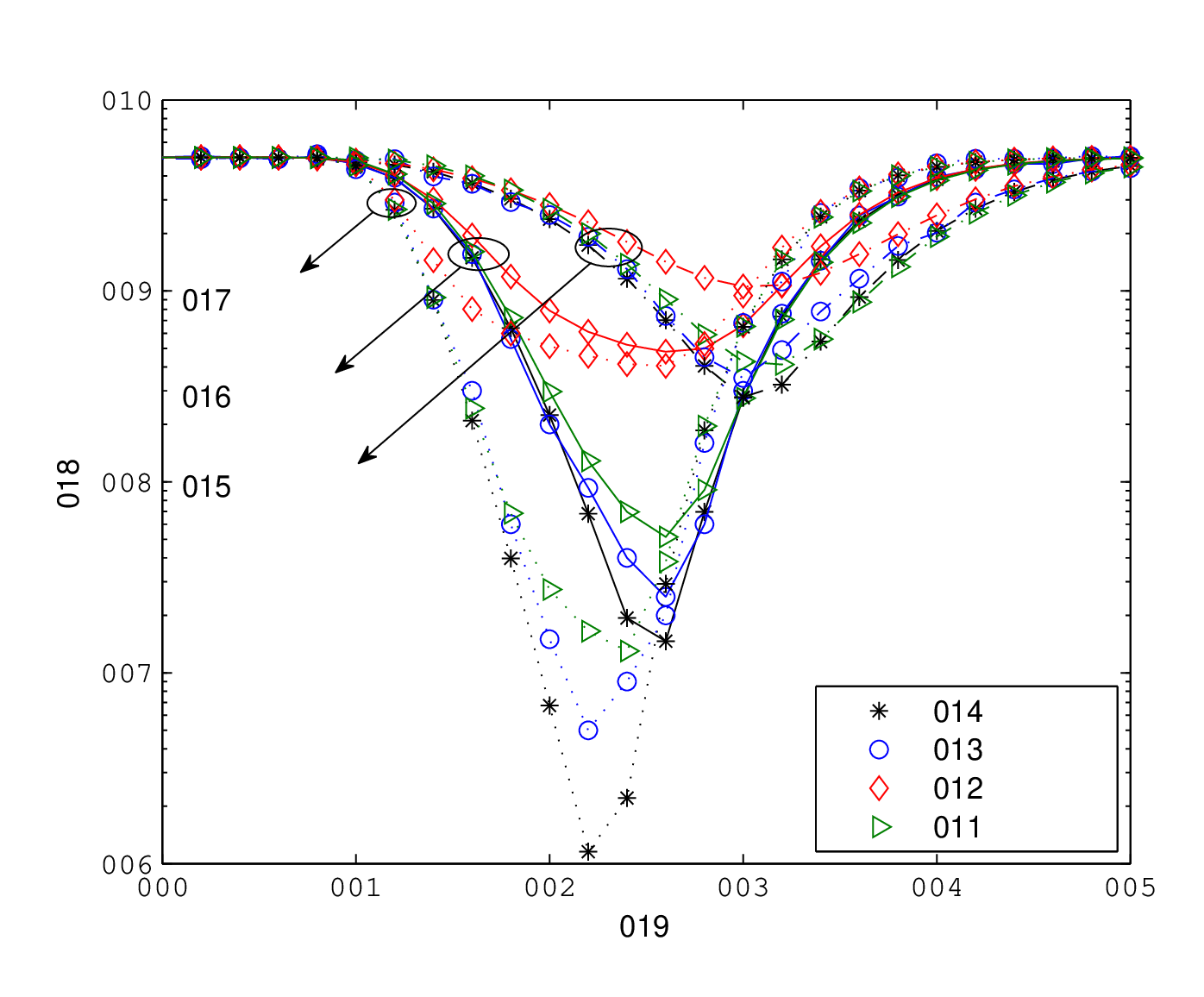}}\vspace{-0.4cm}
\caption{ BER versus detection threshold $\xi_A$ for the peak detector and different average symbol durations $\bar{T}^{\mathrm{symb}}$.  }
\label{Fig:BER_Tsymb}
  \end{minipage}
  \hfill
  \begin{minipage}[b]{0.1\textwidth}
  \end{minipage} \vspace{-0.4cm}
\end{figure*}

Next, we study the performance of the proposed synchronization schemes in terms of the end-to-end BER. First,  the impact of synchronization threshold $\xi_B$ on the proposed  TT synchronization scheme is investigated. To this end, in Fig.~\ref{Fig:BER_xiB}, we show the BERs of the mean and peak detectors versus synchronization threshold $\xi_B$ for different detection thresholds $\boldsymbol{\xi}_A=[\xi_A^{\mathtt{mean}},\xi_A^{\mathtt{peak}}]=[4,12],[4.5,13]$, and $[5,14]$ where $\xi_A^{\mathtt{mean}}$ and $\xi_A^{\mathtt{peak}}$ are the detection thresholds used for the mean and peak detectors, respectively. We observe that for each curve, there  exists an optimal value for the synchronization threshold which minimizes the BER. Moreover, for the considered example, for each detector, the optimal value of $\xi_B$ does not depend on the detection threshold $\xi_A$. In particular, in Fig.~\ref{Fig:BER_xiB}, the optimal values for $\xi_B$ for the mean and peak detectors are $13$ and $12$, respectively, for all considered values of the detection thresholds, i.e., for the peak detector, the optimal $\xi_B$ is smaller than for the mean detector. This may be due to the fact that for the peak detector, the peak value of $r_A(t_n)$ within a symbol interval determines the result of detection, and smaller $\xi_B$ make it more likely that the peak is not missed.

Fig.~\ref{Fig:BER_xiA} shows the BERs of the mean and peak detectors as functions of the detection threshold $\xi_A$. For the proposed TT scheme, we choose $\xi_B = 13$ and $\xi_B=12$  for the mean and peak detectors, respectively. As a general trend for all curves, we observe from Fig.~\ref{Fig:BER_xiA} that the BER is minimized for a specific value (values) of the detection threshold $\xi_A$. Furthermore, since only one observation is employed for the peak detector, the BER depends on $\lceil\xi_A\rceil$. Therefore, for the peak detector, all detection thresholds $i<\xi_A\leq i+1$, where $i$ is an integer number, yield identical BER. We further note that the different schemes in Fig.~\ref{Fig:BER_xiA} should be compared based on their minimum BER, i.e., for that $\xi_A$, where the BER is minimized. We observe from Fig.~\ref{Fig:BER_xiA} that ML synchronization provides a BER performance close to that of perfect synchronization where the starts of the symbol intervals are assumed to be perfectly known. The PO synchronization scheme leads to a considerable BER performance loss compared to ML synchronization. In this regard, the proposed TT synchronization scheme provides a favorable tradeoff between complexity and BER performance.  

Finally, in Fig.~\ref{Fig:BER_Tsymb}, we investigate the effect of the average symbol duration on the performance of the proposed synchronization schemes.  We depict the BER versus the detection threshold $\xi_A$, and for clarity of presentation, only results for the peak detector are shown. Similar results can be obtained for the mean detector.  The ratio $\frac{T^{\max}}{T^{\min}}=1.5$ is kept constant and we consider average symbol durations of $0.5$, $1$, and $2$ ms. Furthermore, the value of the synchronization threshold for the TT scheme is chosen such that the minimum BER for each curve is minimized. This leads to $\xi_B=10, 12$, and $17$ for $\bar{T}^{\mathrm{symb}}=2, 1$, and $0.5$, respectively.  It is observed from Fig.~\ref{Fig:BER_Tsymb} that as the size of the symbol interval decreases, the performances of all considered synchronization schemes, even that for perfect synchronization, deteriorate. This is due to the fact that  as the symbol duration decreases, the ISI increases which degrades not only the performance of the considered synchronization schemes (expect perfect synchronization) but also that of the considered detection scheme.

%

\section{Conclusions}

In this paper, we considered an MC system where the transmitter is not equipped with an internal clock and is not restricted to emit the molecules with a perfect release frequency. To enable symbol synchronization in this case, we proposed to employ two types of molecules, one for synchronization and one for data transmission. We derived the optimal ML synchronization scheme as a performance upper bound. As ML synchronization entails high complexity, we also developed two low-complexity synchronization schemes, namely the suboptimal PO  and TT  schemes. 
In the following, we summarize the main features and drawbacks of the symbol synchronization schemes proposed in this paper. The ML synchronization scheme is the most accurate but also the most complex scheme and requires full knowledge of the MC channel. The PO synchronization scheme is the simplest scheme since it needs the least a priori information about the channel and does not require any additional constraint (other than those introduced in the system model in Section~II); however, our simulation results revealed that it may also introduce a significant BER performance loss.  Compared with the ML and PO synchronization  schemes, the TT synchronization scheme  provides a favorable tradeoff between complexity and BER performance which makes it well suited for applications in practical MC systems with limited computational capabilities.

\bibliographystyle{IEEEtran}
\bibliography{Ref_23_02_2017}

%

\end{document}